\documentclass[twocolumn]{webofc}

\graphicspath{{./}{figs/}}

\usepackage{aas_macros} 
\usepackage[varg]{txfonts}   

\newcommand{\ENDF}{ENDF/B-VIII.0}
\begin{document}
%
\title{Nuclear data activities for medium mass and heavy nuclei at Los Alamos}
%
%

\author{
\firstname{M. R.} \lastname{Mumpower}\inst{1}\fnsep\thanks{e-mail: mumpower@lanl.gov}
\and
\firstname{T. M.} \lastname{Sprouse}\inst{1}
\and
\firstname{T.} \lastname{Kawano}\inst{1}
\and
\firstname{M. W.} \lastname{Herman}\inst{1}
\and
\firstname{A. E.} \lastname{Lovell}\inst{1}
\and
\firstname{G. W.} \lastname{Misch}\inst{1}
\and
\firstname{D.} \lastname{Neudecker}\inst{1}
\and
\firstname{H.} \lastname{Sasaki}\inst{1}
\and
\firstname{I.} \lastname{Stetcu}\inst{1}
\and
\firstname{P.} \lastname{Talou}\inst{1}
}

\institute{Los Alamos National Laboratory, Los Alamos, NM, 87545, USA
}

\abstract{%
Nuclear data is critical for many modern applications from stockpile stewardship to cutting edge scientific research.
Central to these pursuits is a robust pipeline for nuclear modeling as well as data assimilation and dissemination.
We summarize a small portion of the ongoing nuclear data efforts at Los Alamos for medium mass to heavy nuclei.
We begin with an overview of the \texttt{NEXUS} framework and show how one of its modules can be used for model parameter optimization using Bayesian techniques.
The mathematical framework affords the combination of different measured data in determining model parameters and their associated correlations.
It also has the advantage of being able to quantify outliers in data.
We exemplify the power of this procedure by highlighting the recently evaluated $^{239}$Pu cross section.
We further showcase the success of our tools and pipeline by covering the insight gained from incorporating the latest nuclear modeling and data in astrophysical simulations as part of the Fission In R-process Elements (FIRE) collaboration.
}
\maketitle
%
\section{Introduction}
\label{sec:intro}

Nuclear data have a profound impact on modern society \cite{Kolos2022}.
These data serve as a foundation for nuclear energy, nuclear security, and the wide variety of research that is carried out in nuclear astrophysics from the composition of neutron stars to the impact of kilonova light curves \cite{Schatz2016, Bernstein2019, Barnes2021, Schatz2022}.

Despite its importance, a robust pipeline for nuclear data assimilation and dissemination still remains on the horizon.
Currently, major evaluated databases of nuclear reactions, decays and structure can be found in ENDF \cite{ENDF8} and ENSDF \cite{ENSDF}.
Inputs valuable for reaction modeling codes can be found in the Reference Input Parameter Library (RIPL) \cite{Capote2009} and experimental reaction data can be found in EXFOR \cite{Otuka2014}.
The Atomic Mass Evaluation (AME) \cite{Wang2017, Wang2021} and NuBase \cite{Kondev2021} provide targeted information regarding ground state and decay properties.

The disparate nature of these sources, coupled with the difficulty of extracting information, complicate the use of these database in modern applications, particularly with cutting edge science.
In this contribution, we provide a brief overview of the nuclear data activities at Los Alamos National Laboratory (LANL).
We discuss the \texttt{NEXUS} framework which represents a first step towards providing a pipeline between nuclear data, nuclear modeling efforts and applications.
We showcase the utility of this platform by highlighting its use in a recent evaluation of $^{239}$Pu cross sections as well as in the scientific Fission In R-process Elements (FIRE) collaboration.
We end with a discussion of recent tools developed at LANL intended to advance the use of nuclear data in astrophysical applications.

\section{NEXUS}
\label{sec:nexus}

The Los Alamos Python package, \texttt{NEXUS}, is a data-agnostic framework intended to furnish access to nuclear properties.
The phrase `data agnostic' means that it does not matter how the data was generated or parsed.
The output format of the data is also irrelevant to the framework's function.
Instead the focus of the code is on a consistent object-oriented representation of various physical quantities and the relationships between them.
This approach is in stark contrast to the major databases which revolve around the transmission format of data.

As an example of an object in \texttt{NEXUS}, in its simplest form a reaction cross section may be reasonably represented by two arrays: the incident particle energy and the cross section at each energy.
Additional information may be warranted, in which case the base object may be extended.
There may be uncertainties reported on each energy point as well as uncertainties in the cross section values.
For particular applications associated metadata may also need to be affiliated with attributes, for example, providing the physical units of each of the arrays.
The focal point is on the representation of the data in code, not on how the information will be transmitted.

Information transmission of data is supported however in \texttt{NEXUS}.
This revolves around the concept of marshalling which is a more general concept of object serialization (which deals only with string representations).
Each object in \texttt{NEXUS} has a corresponding object, a marshaller, which handles its transmutation to various representations.
A marshaller may translate one object type into another, or it may convert an object from its code representation to a form that can be saved on a computer hard disk, e.g. ASCII or JSON.
The power of this approach means that complex derived objects may be constructed and converted to the desired output format for particular applications.
A concert example is the use of atomic binding energies and their appropriate parsing into suitable reaction network codes for use in astrophysical \cite{Li2022} or Machine Learning applications \cite{Lovell2022, Mumpower2022}.

The \texttt{NEXUS} framework provides a base set of objects for nuclear properties that are commonly found in the aforementioned nuclear databases.
In addition, access to a host of theoretical outputs from nuclear model codes are also available.
Among other data access methods, this list includes, level densities, $\gamma$-strength functions, optical model parameters, ground state binding energies, and reaction cross sections.
Information regarding nuclear isomers has recently been utilized to study the impact of long-lived excited states in astrophysics \cite{Misch2020, Misch2021}.

\begin{figure}[h]
\centering
\includegraphics[width=80mm]{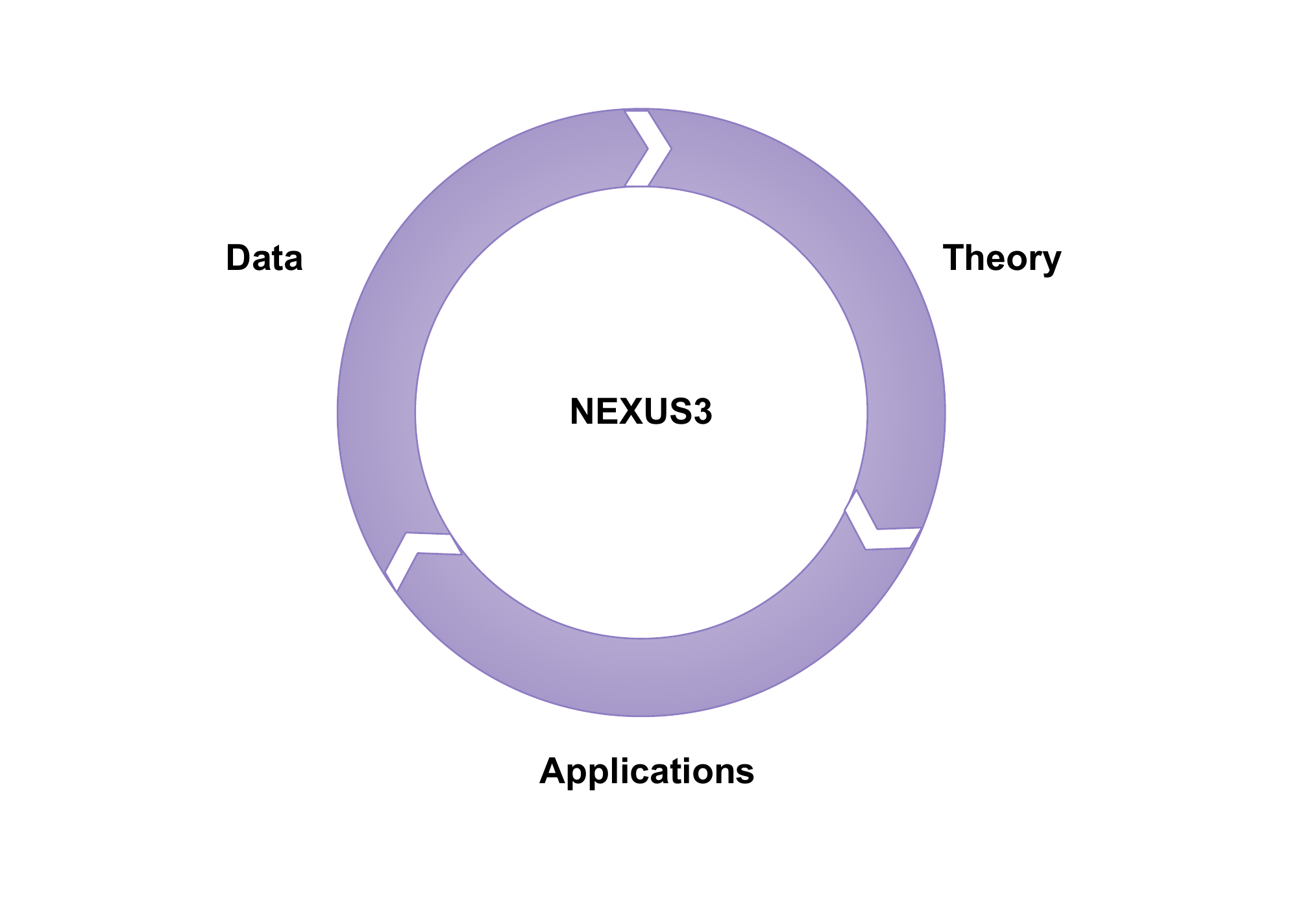}
\caption{The Los Alamos Python package, \texttt{NEXUS}, provides a series of application programming interfaces (APIs) to efficiently go from nuclear theory or data to applications and back again. }
\label{fig:nexus3}
\end{figure}

The data-agnostic approach \texttt{NEXUS} framework also provides a set of application programming interfaces (APIs) to nuclear data libraries, nuclear model codes and application codes.
The transition between these three areas enabled by this platform is depicted pictorially in Figure \ref{fig:nexus3}.

\section{Hierarchical Approach to Nuclear Data Evaluations}
\label{sec:bopt}

We now highlight the power of the \texttt{NEXUS} framework by showcasing a result from a recent evaluation of $^{239}$Pu which uses the Bayesian optimization module.
The evaluation is based on a hierarchical approach where data is deemed most important followed by model calculation to `fill in the gaps' when empirical data is lacking.
This evaluation combines experimental cross section data from EXFOR along with output from the Los Alamos statistical model code, CoH \cite{Kawano2016, Kawano2019}.

In order to maintain consistency throughout each evaluated cross section channel, a global set of model parameters must be fit to available data.
Model parameter fitting is performed incrementally channel by channel starting with the total cross section.
Because most statistical model cross sections perform very well with respect to available data, it is possible to use a Bayesian approach called hyperparameter optimization to optimize the model with respect to independent datasets \cite{Hobson2002}.

To optimize the calculated total cross section with respect to available data, $\sigma_\textrm{tot}$, a Metropolis random-walk algorithm is used to probe the optical model parameter space.
The Soukhovitskii potential is used as the basis for the optical model \cite{Soukhovitskii2005, Capote2005}.
The model space consists of six parameters, including the potential depths, diffuseness and radii.
All other model parameters are held fixed during this optimization as they do not influence the total cross section.
It was determined that holding the optical model deformation fixed, rather than letting it vary during the optimization was ideal for approximating a robust minimum.
In this procedure the chi-square goodness of fit is minimized with respect to data which is weighted based on its uncertainty.

\begin{figure}[h]
\centering
\includegraphics[width=80mm]{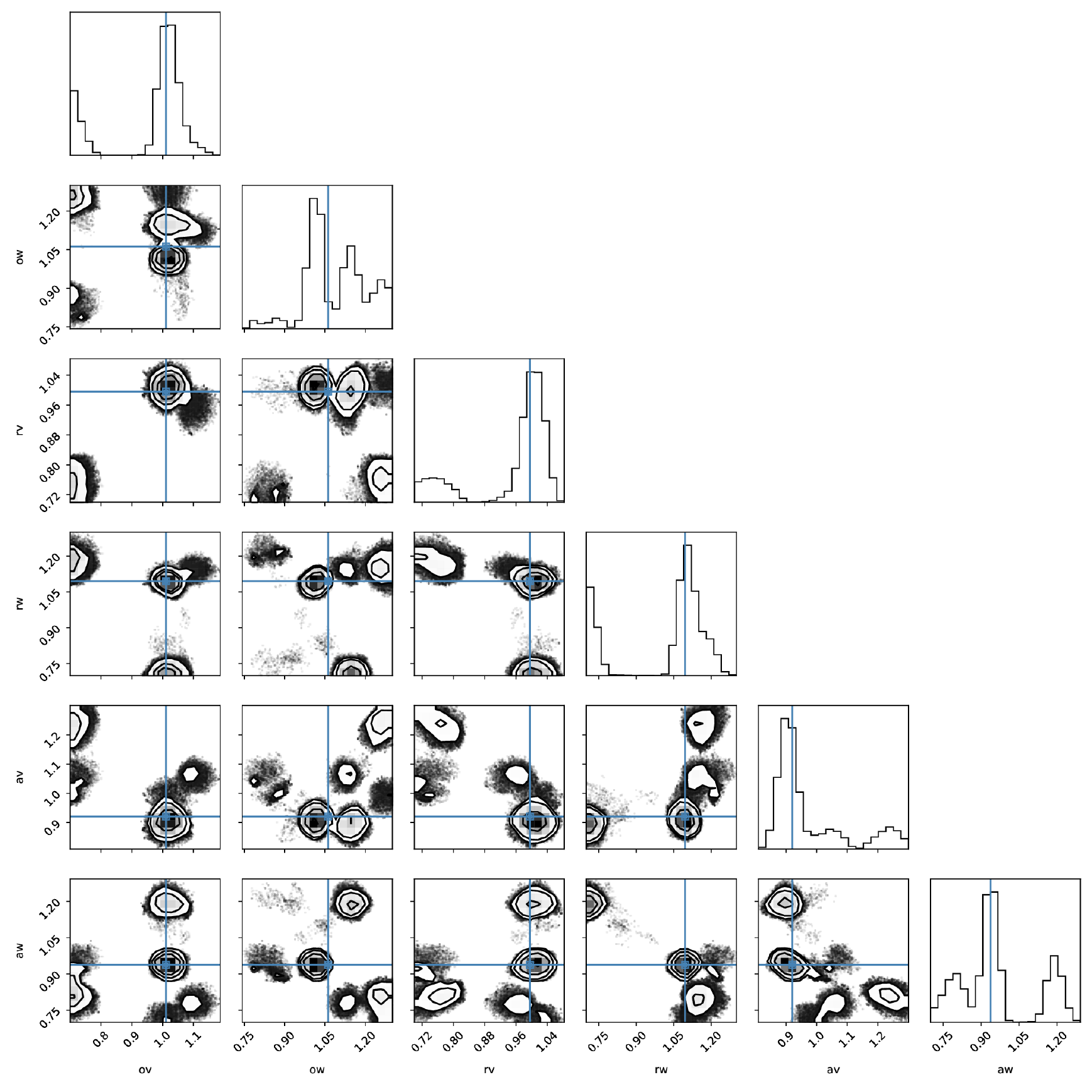}
\caption{The determination of optical model parameters using Bayesian hyperparameter optimization. Shaded regions highlight areas of minimal $\chi^2$. }
\label{fig:cst_bopt}
\end{figure}

The resultant parameter sensitivities of the fitting procedure are shown in Figure \ref{fig:cst_bopt}.
The optical model parameters are standardized to unity to ensure that there is no difference in the scale of individual parameters.
All of the model parameters are found to exhibit a relatively strongly peaked distribution around the optimal value.
The optimal values are all within 10\% from the baseline parameters suggested by Soukhovitskii with diffuseness parameters changing the most.

\begin{figure}[h]
\centering
\includegraphics[width=80mm]{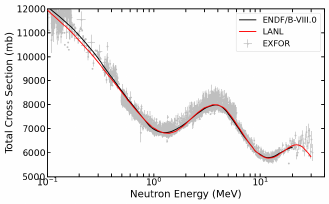}
\caption{The evaluation of the total cross section of neutrons on $^{239}$Pu using the Los Alamos code \texttt{CoH} for cross section modeling. }
\label{fig:cst_pu}
\end{figure}

The cross section fit with the optimal optical model parameters are shown in Fig.~\ref{fig:cst_pu}.
The procedure performs nearly identically to the previous evaluation of \ENDF{}.
Slight modifications are seen relative to \ENDF{} below 30 keV where the reported uncertainties of the datasets pull down the fit to the total cross section.
A similar modification, albeit to a much smaller effect, can be found between 1 and 2 MeV, near the local minimum of $\sigma_\textrm{tot}$.

\section{FIRE Collaboration}

The \texttt{NEXUS} framework also serves as the basis for the nuclear data pipeline used in the highly successful Fission In R-process Elements (FIRE) topical collaboration.
While the name centers around fission, many relevant nuclear properties for this nucleosynthesis process have been studied by FIRE researchers \cite{Orford2018, Vilen2018, Vilen2020, Wang2020a, Wang2020b, Cote2021, Zhu2022}.
Nuclear data appropriate for studies of nuclear astrophysics is assimilated and disseminated via a tailored repository which transform data into the appropriate format for use in the Los Alamos Portable Routines for Integrated nucleoSynthesis Modeling (PRISM) reaction network code \cite{Sprouse2021}.
This state-of-the-art nuclear reaction network has been frequently used in contemporary studies from the study of universality among metal poor halo stars \cite{Roederer2022} to the first kilonova actinide opacity results \cite{Fontes2022}.
We now cover several recent FIRE studies that highlight the crucial role accurate nuclear data can play in understanding the formation of the heaviest elements in the cosmos.

The synthesis of the heaviest elements in nature is thought to occur via the rapid neutron capture ($r$-) process \cite{Kajino2019}.
While the astrophysical event(s) responsible for this process are still under debate \cite{Thielemann2011}, it is with certainty that nuclear properties (most influentially: masses, half-lives, reaction rates, and fission properties) are critical for understanding this phenomenon \cite{Mumpower2016}.
The central problem to this endeavor is that accurate data is limited to a small portion of the chart of nuclides and therefore simulations of the $r$ process must rely on theoretical extrapolations for thousands of nuclei when this data is not available \cite{Sprouse2020}.
This implies that to understand $r$-process nucleosynthesis, it is of paramount importance to simultaneously include the most accurate nuclear data and modern theoretical predictions.

Once the $r$ process creates heavy, radioactive elements, they must decay back to more stable isotopes.
In doing so, the radioactive decay of these freshly synthesized elements power a distinct light curve in compact object merger  environments (neutron star mergers and black hole neutron star mergers) called a kilonova.
The importance of $\beta$-decay half-lives and branching ratios on the impact of kilonova light curves has been studied in the recent work of Lund \textit{et al.} (2022) \cite{Lund2022}.
Lund and colleagues showed that different theoretical models of $\beta$-decay properties profoundly impact the production of elements that undergo $\alpha$-decay and spontaneous fission.
The influence of $\alpha$ and spontaneous fission emitters in turn strongly impacts nuclear heating that powers the light curves.
Neutron-rich isotopes of nobelium ($Z=102$), lawrencium ($Z=103$), rutherfordium ($Z=104$), and dubnium ($Z=105$) were found to be particularly influential on observational timescales.
This lead the researchers to conclude that the $\beta$-decay feeders of these nuclei represent a prime target for future precision experimental campaigns.
More comprehensive studies of the impact of nuclear data and model uncertainties on kilonova were performed in Refs.~\cite{Zhu2021, Barnes2021}.

The inclusion of measured $\beta$-decay properties has also been studied in the context of this pipeline.
Kiss \textit{et al.} measured 28 new $\beta$-decay neutron emission probabilities and half-lives for important lanthanide species \cite{Kiss2022}.
This study not only informed the impact of measurements in astrophysical environments, it was able to rank the importance of the measurements through novel Bayesian analysis.
Using a new measure to gauge the impact of each measurement, the final abundances were shown to be most sensitive to the half-life of $^{168}$Sm followed by several Gd isotopes.
This work is in agreement with another experimental effort supported by FIRE scientists studying $\beta$-delayed neutron emitters that were found to be influential around the second or $A \sim 130$ $r$-process peak \cite{Hall2021}.

FIRE researchers have also teamed up with nuclear experimentalists to study nuclear structure measurements \cite{Orford2022} including the first exploration of neutron shell structure to the `south east' of lead \cite{Tang2020}.
The production of $^{207}$Hg led the team to conclude that current data suggests a strong shell closure at neutron number $N=126$, in agreement with modern theoretical model predictions.
The $N=126$ is the final confirmed closed shell in the chart of nuclides.
The strength of this shell closure acts as a `gatekeeper' for the production of actinides and is thus extremely important in their production as well as for how much fission may ensue in an $r$-process event.

In one of the major results of the FIRE collaboration, fission fragment mass yields were calculated for all fissioning systems using the Finite-Range Liquid-Drop Model (FRLDM) \cite{Mumpower2020}.
The statistics of each fissioning system was carefully compiled using millions of computational hours on the Los Alamos high-performance computing network.
One of the fundamental observations made by the team was that the spread of the fission yield distribution increases as a function of neutron excess.
As a consequence of this result, the heaviest neutron-rich nuclei that undergo fission in the $r$ process would spread their fragments over a much larger mass range than previous studies anticipated.
The impact of these yields were subsequently tested in the astrophysical $r$ process by Vassh \textit{et al.} (2020).
Vassh and colleagues found that the wide fission yields of FRLDM drastically improve the fit to the solar isotopic residuals relative to the commonly used symmetric split technique.
Wide fission yields impact the production of lanthanides as well as the production of lighter precious metals.
The possibility of co-production among these elements is in agreement with observations of metal poor halo stars which are thought to be evident of a single $r$-process event.

\section{URSA}

With the dramatic and ongoing acceleration in the  technical capabilities offered by recent and upcoming nuclear experimental facilities and detectors, such as FRIB \cite{Horowitz2019}, N=126 Factory \cite{Savard2020}, and storage rings as GSI, IMP, and RIKEN \cite{Zhang2016}, there will be a rapid transformation in state-of-the-art nuclear data made available to astrophysical applications, e.g., in astrophysical nucleosynthesis \cite{Rauscher2002,Kajino2019, Cowan2021}, stellar structure and evolution \cite{Paxton2011}, and galactic chemical evolution \cite{Cote2018}, among other possibilities.
Over the past 50 years \cite{B2FH}, the intimate relationship between nuclear physics, astrophysics, and astronomy has become increasingly interconnected.
As such, we are at a critical point where the need to bridge the connections between the nuclear physics and astrophysics communities has never been more apparent \cite{Meynet2008, Arnould2020}.

\begin{figure}[h]
\centering
\includegraphics[width=80mm]{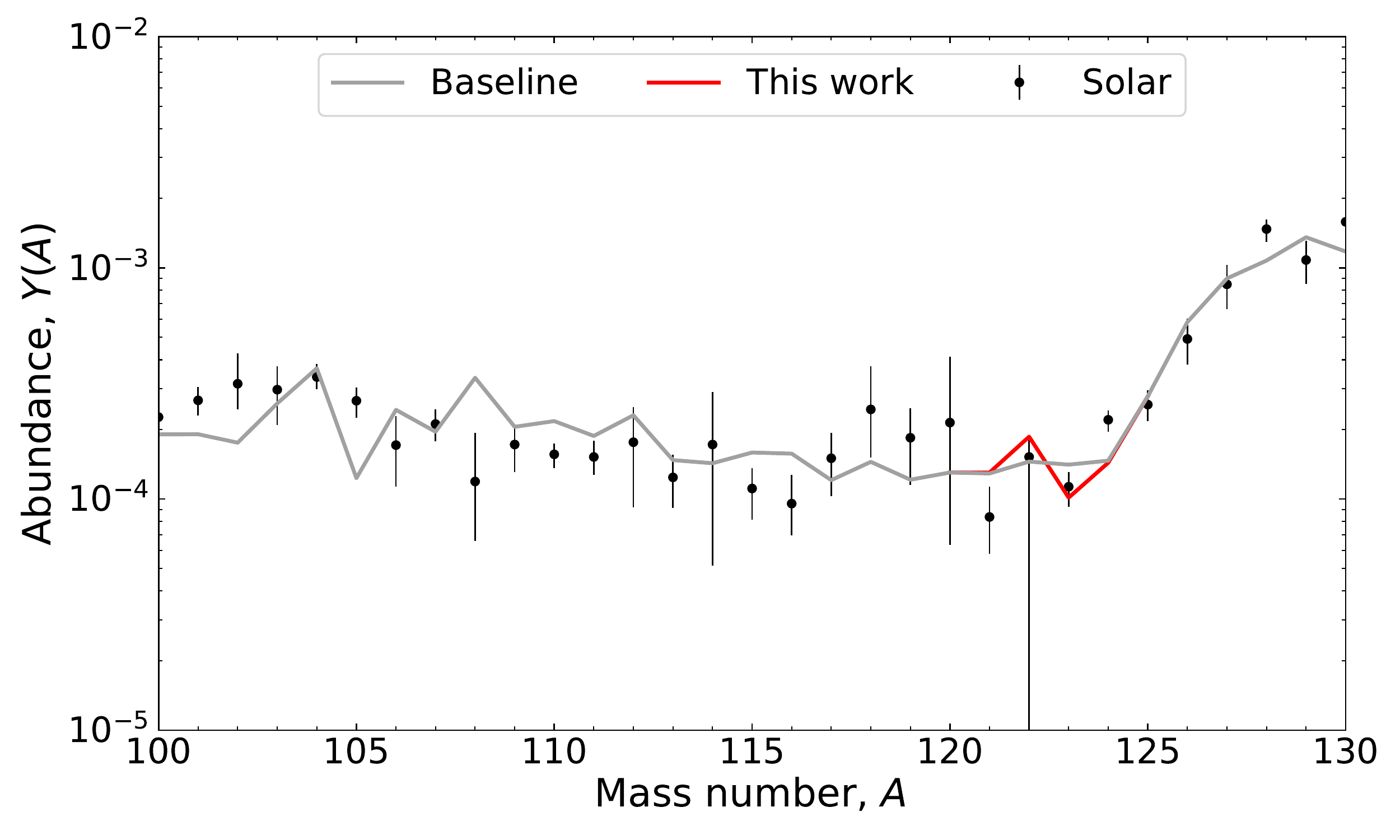}
\caption{The impact on the final $r$-process abundances from the single measurement of the mass of $^{123}$Pd (red curve) relative to the baseline simulation \cite{Li2022}. The $r$-process solar isotopic residuals are shown as black dots. This result marked the first time a superposition of trajectories were used to study the impact of a new measurement, in contrast to present day sensitivity studies which use only single trajectories.  }
\label{fig:rp_li}
\end{figure}

The rapid neutron capture process, or $r$ process, is an especially prominent example of the connection between astrophysics and nuclear physics: the $r$ process proceeds through some of the most neutron-rich nuclei thought to exist in the universe, which establishes this unique category of astrophysical environments as an excellent probe of the nuclear physics of nuclei far away from the valley of stability and approaching the neutron dripline \cite{Quark2Cosmos}.
At the same time, this makes the $r$ process extraordinarily sensitive to nuclear physics uncertainties which currently plague these same neutron-rich nuclei \cite{Mumpower2016}, highlighting the need within the computational astrophysics community to incorporate even individual points of new experimental nuclear data from the moment it becomes available.
As an example of this effect, we refer to Fig.~\ref{fig:rp_li}, where a single mass measurement is seen to adjust the $r$ process abundance pattern by a relatively significant amount \cite{Li2022}; with the advent of many ($\approx$ 100s)  of similarly impactful experiments expected to occur in the immediate- and near-future \cite{Surman2018}, it is reasonable to expect that the astrophysics community is facing a wonderful opportunity to capitalize on the ongoing developments in nuclear physics and nuclear data communities.

\begin{figure*}[h]
\centering
\includegraphics[width=160mm]{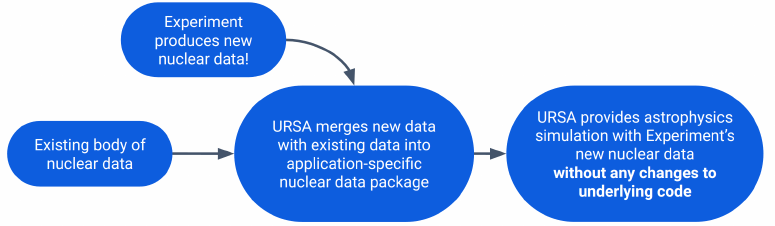}
\caption{An example of the Unified Reaction Structures for Astrophysics (\texttt{URSA}) data-to-application pipeline. From the moment that even a single point of nuclear data is made available, \texttt{URSA} provides a complete suite of tools to update data in existing files within its prescribed format without any need to modify existing simulation codebases.}
\label{fig:ursa}
\end{figure*}

In anticipation of these advances, we have begun development of Unified Reaction Structures in Astrophysics (\texttt{URSA}), which aims to expedite the transmission of nuclear data from first experiment to computational simulations in astrophysics.
The overall motivation of \texttt{URSA} is centered around the complete decoupling of computational applications of nuclear data from specific details regarding the data formats and/or data sources themselves.
In practical terms, this manifests as an underlying data format that is sufficiently expressive to capture the many types of nuclear data needs in astrophysical and related applications, which variously include radiation decay spectra, reaction cross sections, decay half-lives and branching ratios, fission properties, among numerous other conceivable possibilities.
At the same time, \texttt{URSA} intends to provide, alongside the data format itself, numerical libraries that access, process, and interpret data in any such ways that numerical simulation codes might require.
In this way, a computational astrophysicist need not rely on other researchers or evaluators to import any particular piece of experimental or evaluated data into one data format or another; rather, \texttt{URSA} provides the resources to update all aspects of computational simulations' nuclear data requirements, as necessary, to incorporate new nuclear data \textit{in real time} as it becomes available.
The ideal data-to-application pipeline suggested by this model is demonstrated schematically by Fig.~\ref{fig:ursa}, where we outline the efficiency with which \texttt{URSA} can enable new and innovative nuclear data to be implemented in astrophysical applications.
The overall effect is to enable an immediate reaction time to developments in nuclear data across a broad swath of astrophysical nuclear data applications.
Two factors that will be critical as  the \texttt{URSA} project evolves will be community uptake and user feedback, without either of which the full potential of its impacts will surely not be realized.

\section{Summary}

Access and incorporation of accurate nuclear data is influential to producing reliable nuclear evaluations and empowers scientific discovery.
We have reviewed a handful of cases where the Los Alamos \texttt{NEXUS} framework along with theoretical model codes have been used in this regard.
With a stream of new nuclear data expected to be produced by radioactive beam facilities such as at the Facility for Radioactive Isotope Beams (FRIB), adoption of transmission protocols such as the Unified Reaction Structures for Astrophysics (\texttt{URSA}) will be crucial for researchers in order to keep pace with rapidly evolving scientific advancement. 

\section*{Acknowledgements}

M.R.M. would like to acknowledge valuable discussions with Mark Chadwick, Kay Kolos, Filip Kondev, Michael Smith and Patrick Talou.
The authors were supported by the US Department of Energy through the Los Alamos National Laboratory (LANL).
LANL is operated by Triad National Security, LLC, for the National Nuclear Security Administration of U.S.\ Department of Energy (Contract No.\ 89233218CNA000001).
This work is intended for unlimited release with identification number: LA-UR-22-30993.

\bibliography{refs}

\begin{thebibliography}{55}

\bibitem{Kolos2022}
K.~Kolos, V.~Sobes, R.~Vogt, C.E. Romano, M.S. Smith, L.A. Bernstein, D.A.
  Brown, M.T. Burkey, Y.~Danon, M.A. Elsawi et~al., Phys. Rev. Research
  \textbf{4}, 021001 (2022)

\bibitem{Schatz2016}
H.~Schatz, Journal of Physics G: Nuclear and Particle Physics \textbf{43},
  064001 (2016)

\bibitem{Bernstein2019}
L.A. Bernstein, D.A. Brown, A.J. Koning, B.T. Rearden, C.E. Romano, A.A.
  Sonzogni, A.S. Voyles, W.~Younes, Annual Review of Nuclear and Particle
  Science \textbf{69}, 109 (2019),
  \texttt{https://doi.org/10.1146/annurev-nucl-101918-023708}

\bibitem{Barnes2021}
J.~{Barnes}, Y.L. {Zhu}, K.A. {Lund}, T.M. {Sprouse}, N.~{Vassh}, G.C.
  {McLaughlin}, M.R. {Mumpower}, R.~{Surman}, \apj \textbf{918}, 44 (2021),
  \texttt{2010.11182}

\bibitem{Schatz2022}
H.~Schatz, A.D.B. Reyes, A.~Best, E.F. Brown, K.~Chatziioannou, K.A. Chipps,
  C.M. Deibel, R.~Ezzeddine, D.K. Galloway, C.J. Hansen et~al., \emph{Horizons:
  Nuclear astrophysics in the 2020s and beyond} (2022),
  \urlstyle{tt}\url{https://arxiv.org/abs/2205.07996}

\bibitem{ENDF8}
D.A. Brown, M.B. Chadwick, R.~Capote, A.C. Kahler, A.~Trkov, M.W. Herman, A.A.
  Sonzogni, Y.~Danon, A.D. Carlson, M.~Dunn et~al., Nuclear Data Sheets
  \textbf{148}, 1  (2018)

\bibitem{ENSDF}
\emph{Ensdf database}, \url{http://www.nndc.bnl.gov/ensarchivals/}, accessed:
  2022-10-14

\bibitem{Capote2009}
R.~Capote, M.~Herman, P.~Obložinský, P.~Young, S.~Goriely, T.~Belgya,
  A.~Ignatyuk, A.~Koning, S.~Hilaire, V.~Plujko et~al., Nuclear Data Sheets
  \textbf{110}, 3107 (2009), special Issue on Nuclear Reaction Data

\bibitem{Otuka2014}
N.~Otuka, E.~Dupont, V.~Semkova, B.~Pritychenko, A.~Blokhin, M.~Aikawa,
  S.~Babykina, M.~Bossant, G.~Chen, S.~Dunaeva et~al., Nuclear Data Sheets
  \textbf{120}, 272 (2014)

\bibitem{Wang2017}
M.~Wang, G.~Audi, F.G. Kondev, W.~Huang, S.~Naimi, X.~Xu, Chinese Physics C
  \textbf{41}, 030003 (2017)

\bibitem{Wang2021}
M.~Wang, W.~Huang, F.~Kondev, G.~Audi, S.~Naimi, Chinese Physics C \textbf{45},
  030003 (2021)

\bibitem{Kondev2021}
F.~Kondev, M.~Wang, W.~Huang, S.~Naimi, G.~Audi, Chinese Physics C \textbf{45},
  030001 (2021)

\bibitem{Li2022}
H.F. {Li}, S.~{Naimi}, T.M. {Sprouse}, M.R. {Mumpower}, Y.~{Abe},
  Y.~{Yamaguchi}, D.~{Nagae}, F.~{Suzaki}, M.~{Wakasugi}, H.~{Arakawa} et~al.,
  \prl \textbf{128}, 152701 (2022), \texttt{2112.05312}

\bibitem{Lovell2022}
A.E. Lovell, A.T. Mohan, T.M. Sprouse, M.R. Mumpower, Phys. Rev. C
  \textbf{106}, 014305 (2022)

\bibitem{Mumpower2022}
M.R. Mumpower, T.M. Sprouse, A.E. Lovell, A.T. Mohan, Phys. Rev. C
  \textbf{106}, L021301 (2022)

\bibitem{Misch2020}
G.W. Misch, S.K. Ghorui, P.~Banerjee, Y.~Sun, M.R. Mumpower, The Astrophysical
  Journal Supplement Series \textbf{252}, 2 (2020)

\bibitem{Misch2021}
G.W. Misch, T.M. Sprouse, M.R. Mumpower, The Astrophysical Journal Letters
  \textbf{913}, L2 (2021)

\bibitem{Kawano2016}
T.~Kawano, R.~Capote, S.~Hilaire, P.~Chau Huu-Tai, Phys. Rev. C \textbf{94},
  014612 (2016)

\bibitem{Kawano2019}
T.~Kawano, \emph{Unified coupled-channels and hauser-feshbach model calculation
  for nuclear data evaluation} (2019),
  \urlstyle{tt}\url{https://arxiv.org/abs/1901.05641}

\bibitem{Hobson2002}
M.P. {Hobson}, S.L. {Bridle}, O.~{Lahav}, \mnras \textbf{335}, 377 (2002),
  \texttt{astro-ph/0203259}

\bibitem{Soukhovitskii2005}
E.S. Soukhovitskii, R.~Capote, J.M. Quesada, S.~Chiba, Phys. Rev. C
  \textbf{72}, 024604 (2005)

\bibitem{Capote2005}
R.~Capote, E.S. Soukhovitskii, J.M. Quesada, S.~Chiba, Phys. Rev. C
  \textbf{72}, 064610 (2005)

\bibitem{Orford2018}
R.~{Orford}, N.~{Vassh}, J.A. {Clark}, G.C. {McLaughlin}, M.R. {Mumpower},
  G.~{Savard}, R.~{Surman}, A.~{Aprahamian}, F.~{Buchinger}, M.T. {Burkey}
  et~al., \prl \textbf{120}, 262702 (2018)

\bibitem{Vilen2018}
M.~{Vilen}, J.M. {Kelly}, A.~{Kankainen}, M.~{Brodeur}, A.~{Aprahamian},
  L.~{Canete}, T.~{Eronen}, A.~{Jokinen}, T.~{Kuta}, I.D. {Moore} et~al., \prl
  \textbf{120}, 262701 (2018), \texttt{1801.08940}

\bibitem{Vilen2020}
M.~{Vilen}, J.M. {Kelly}, A.~{Kankainen}, M.~{Brodeur}, A.~{Aprahamian},
  L.~{Canete}, R.P. {de Groote}, A.~{de Roubin}, T.~{Eronen}, A.~{Jokinen}
  et~al., \prc \textbf{101}, 034312 (2020), \texttt{1908.05043}

\bibitem{Wang2020a}
X.~{Wang}, {N3AS Collaboration}, B.D. {Fields}, M.~{Mumpower}, T.~{Sprouse},
  R.~{Surman}, N.~{Vassh}, \apj \textbf{893}, 92 (2020), \texttt{1909.12889}

\bibitem{Wang2020b}
X.~{Wang}, {N3AS Collaboration}, N.~{Vassh}, {FIRE Collaboration},
  T.~{Sprouse}, M.~{Mumpower}, R.~{Vogt}, J.~{Randrup}, R.~{Surman}, \apjl
  \textbf{903}, L3 (2020), \texttt{2008.03335}

\bibitem{Cote2021}
B.~{C{\^o}t{\'e}}, M.~{Eichler}, A.~{Yag{\"u}e L{\'o}pez}, N.~{Vassh}, M.R.
  {Mumpower}, B.~{Vil{\'a}gos}, B.~{So{\'o}s}, A.~{Arcones}, T.M. {Sprouse},
  R.~{Surman} et~al., Science \textbf{371}, 945 (2021), \texttt{2006.04833}

\bibitem{Zhu2022}
Y.~{Zhu}, J.~{Barnes}, K.A. {Lund}, T.M. {Sprouse}, N.~{Vassh}, G.C.
  {McLaughlin}, M.R. {Mumpower}, R.~{Surman}, \emph{{The Impact of Nuclear
  Physics Uncertainties on Interpreting Kilonova Light Curves}}, in
  \emph{European Physical Journal Web of Conferences} (2022), Vol. 260 of
  \emph{European Physical Journal Web of Conferences}, p. 03004

\bibitem{Sprouse2021}
T.M. {Sprouse}, M.R. {Mumpower}, R.~{Surman}, \prc \textbf{104}, 015803 (2021),
  \texttt{2008.06075}

\bibitem{Roederer2022}
I.U. {Roederer}, J.J. {Cowan}, M.~{Pignatari}, T.C. {Beers}, E.A. {Den Hartog},
  R.~{Ezzeddine}, A.~{Frebel}, T.T. {Hansen}, E.M. {Holmbeck}, M.R. {Mumpower}
  et~al., \apj \textbf{936}, 84 (2022)

\bibitem{Fontes2022}
C.J. {Fontes}, C.L. {Fryer}, R.T. {Wollaeger}, M.R. {Mumpower}, T.M. {Sprouse},
  \mnras  (2022), \texttt{2209.12759}

\bibitem{Kajino2019}
T.~{Kajino}, W.~{Aoki}, A.B. {Balantekin}, R.~{Diehl}, M.A. {Famiano}, G.J.
  {Mathews}, Progress in Particle and Nuclear Physics \textbf{107}, 109 (2019),
  \texttt{1906.05002}

\bibitem{Thielemann2011}
F.K. Thielemann, A.~Arcones, R.~Käppeli, M.~Liebendörfer, T.~Rauscher,
  C.~Winteler, C.~Fröhlich, I.~Dillmann, T.~Fischer, G.~Martinez-Pinedo
  et~al., Progress in Particle and Nuclear Physics \textbf{66}, 346 (2011),
  particle and Nuclear Astrophysics

\bibitem{Mumpower2016}
M.R. {Mumpower}, R.~{Surman}, G.C. {McLaughlin}, A.~{Aprahamian}, Progress in
  Particle and Nuclear Physics \textbf{86}, 86 (2016), \texttt{1508.07352}

\bibitem{Sprouse2020}
T.M. {Sprouse}, R.~{Navarro Perez}, R.~{Surman}, M.R. {Mumpower}, G.C.
  {McLaughlin}, N.~{Schunck}, \prc \textbf{101}, 055803 (2020),
  \texttt{1901.10337}

\bibitem{Lund2022}
K.A. {Lund}, J.~{Engel}, G.C. {McLaughlin}, M.R. {Mumpower}, E.M. {Ney},
  R.~{Surman}, arXiv e-prints arXiv:2208.06373 (2022), \texttt{2208.06373}

\bibitem{Zhu2021}
Y.L. {Zhu}, K.A. {Lund}, J.~{Barnes}, T.M. {Sprouse}, N.~{Vassh}, G.C.
  {McLaughlin}, M.R. {Mumpower}, R.~{Surman}, \apj \textbf{906}, 94 (2021),
  \texttt{2010.03668}

\bibitem{Kiss2022}
G.G. {Kiss}, A.~{Vit{\'e}z-Sveiczer}, Y.~{Saito}, A.~{Tarife{\~n}o-Saldivia},
  M.~{Pallas}, J.L. {Tain}, I.~{Dillmann}, J.~{Agramunt}, A.~{Algora},
  C.~{Domingo-Pardo} et~al., \apj \textbf{936}, 107 (2022)

\bibitem{Hall2021}
O.~{Hall}, T.~{Davinson}, A.~{Estrade}, J.~{Liu}, G.~{Lorusso}, F.~{Montes},
  S.~{Nishimura}, V.H. {Phong}, P.J. {Woods}, J.~{Agramunt} et~al., Physics
  Letters B \textbf{816}, 136266 (2021)

\bibitem{Orford2022}
R.~{Orford}, N.~{Vassh}, J.A. {Clark}, G.C. {McLaughlin}, M.R. {Mumpower},
  D.~{Ray}, G.~{Savard}, R.~{Surman}, F.~{Buchinger}, D.P. {Burdette} et~al.,
  \prc \textbf{105}, L052802 (2022)

\bibitem{Tang2020}
T.L. Tang, B.P. Kay, C.R. Hoffman, J.P. Schiffer, D.K. Sharp, L.P. Gaffney,
  S.J. Freeman, M.R. Mumpower, A.~Arokiaraj, E.F. Baader et~al., Phys. Rev.
  Lett. \textbf{124}, 062502 (2020)

\bibitem{Mumpower2020}
M.R. {Mumpower}, P.~{Jaffke}, M.~{Verriere}, J.~{Randrup}, \prc \textbf{101},
  054607 (2020), \texttt{1911.06344}

\bibitem{Horowitz2019}
C.J. {Horowitz}, A.~{Arcones}, B.~{C{\^o}t{\'e}}, I.~{Dillmann},
  W.~{Nazarewicz}, I.U. {Roederer}, H.~{Schatz}, A.~{Aprahamian},
  D.~{Atanasov}, A.~{Bauswein} et~al., Journal of Physics G Nuclear Physics
  \textbf{46}, 083001 (2019), \texttt{1805.04637}

\bibitem{Savard2020}
G.~{Savard}, M.~{Brodeur}, J.A. {Clark}, R.A. {Knaack}, A.A. {Valverde},
  Nuclear Instruments and Methods in Physics Research B \textbf{463}, 258
  (2020)

\bibitem{Zhang2016}
Y.H. {Zhang}, Y.A. {Litvinov}, T.~{Uesaka}, H.S. {Xu}, \physscr \textbf{91},
  073002 (2016), \texttt{1811.12003}

\bibitem{Rauscher2002}
T.~{Rauscher}, A.~{Heger}, R.D. {Hoffman}, S.E. {Woosley}, \apj \textbf{576},
  323 (2002), \texttt{astro-ph/0112478}

\bibitem{Cowan2021}
J.J. {Cowan}, C.~{Sneden}, J.E. {Lawler}, A.~{Aprahamian}, M.~{Wiescher},
  K.~{Langanke}, G.~{Mart{\'\i}nez-Pinedo}, F.K. {Thielemann}, Reviews of
  Modern Physics \textbf{93}, 015002 (2021), \texttt{1901.01410}

\bibitem{Paxton2011}
B.~{Paxton}, L.~{Bildsten}, A.~{Dotter}, F.~{Herwig}, P.~{Lesaffre},
  F.~{Timmes}, \apjs \textbf{192}, 3 (2011), \texttt{1009.1622}

\bibitem{Cote2018}
B.~{C{\^o}t{\'e}}, C.L. {Fryer}, K.~{Belczynski}, O.~{Korobkin},
  M.~{Chru{\'s}li{\'n}ska}, N.~{Vassh}, M.R. {Mumpower}, J.~{Lippuner}, T.M.
  {Sprouse}, R.~{Surman} et~al., \apj \textbf{855}, 99 (2018),
  \texttt{1710.05875}

\bibitem{B2FH}
E.M. {Burbidge}, G.R. {Burbidge}, W.A. {Fowler}, F.~{Hoyle}, Reviews of Modern
  Physics \textbf{29}, 547 (1957)

\bibitem{Meynet2008}
G.~{Meynet}, \emph{{Nucleosynthesis from massive stars 50 years after
  B$^{2}$FH}}, in \emph{EAS Publications Series}, edited by C.~{Charbonnel},
  J.P. {Zahn} (2008), Vol.~32 of \emph{EAS Publications Series}, pp. 187--232,
  \texttt{0708.3185}

\bibitem{Arnould2020}
M.~{Arnould}, S.~{Goriely}, Progress in Particle and Nuclear Physics
  \textbf{112}, 103766 (2020), \texttt{2001.11228}

\bibitem{Quark2Cosmos}
B.O.P. {Committee On The Physics Of The Universe}, D.O.E. {Astronomy},
  N.R.C.O.T.N.A. {Physical Sciences}, \emph{{Connecting quarks with the cosmos
  : eleven science questions for the new century}} (2003)

\bibitem{Surman2018}
R.~{Surman}, M.~{Mumpower}, \emph{{Masses and lifetimes for r-process
  nucleosynthesis: FRIB outlook}}, in \emph{European Physical Journal Web of
  Conferences} (2018), Vol. 178 of \emph{European Physical Journal Web of
  Conferences}, p. 04002

\end{thebibliography}

\end{document}